\documentclass[fleqn,twoside]{article}
\usepackage{espcrc2}

\usepackage{graphicx}
\newcommand{\AmS}{{\protect\the\textfont2
  A\kern-.1667em\lower.5ex\hbox{M}\kern-.125emS}}

\newcommand{\eqn}[1]{eqn.~(\ref{#1})}
\title{Baryon operators and baryon spectroscopy}

\author{S.~Basak\address[UMD]{Department of Physics, 
             University of Maryland,
             College Park, MD 20742, USA},
R.G.~Edwards\address[JLAB]{Thomas Jefferson National Accelerator Facility,
             Newport News, VA 23606, USA},
G.T.~Fleming\address[YALE]{Sloane Physics Laboratory, 
             Yale University, 
             New Haven, CT 06520, USA},
U.M.~Heller\address[APS]{American Physical Society,
             Ridge, NY 11961-9000, USA},
A.~Lichtl\address[CMU]{Department of Physics, 
             Carnegie Mellon University, 
             Pittsburgh, PA 15213, USA},
C.~Morningstar\addressmark[CMU],
D.G.~Richards\addressmark[JLAB],
I.~Sato\addressmark[UMD]\address[LBL]{Present address: Nuclear Science Division
Lawrence Berkeley Laboratory
1 Cyclotron Road, MS:70R0319
Berkeley, CA 94720, USA}
S.~Wallace\addressmark[UMD]}

\begin{document}
\begin{abstract}
The issues involved in a determination of the baryon resonance
spectrum in lattice QCD are discussed.  The variational method is
introduced and the need to construct a sufficient basis of
interpolating operators is emphasised.  The construction of baryon
operators using group-theory techniques is outlined.  We find that the
use both of quark-field smearing and link-field smearing in the
operators is essential firstly to reduce the coupling of operators to
high-frequency modes and secondly to reduce the gauge-field
fluctuations in correlators.  We conclude with a status report of our
current investigation of baryon spectroscopy.
\end{abstract}

\maketitle
\section{INTRODUCTION}
Spectroscopy is a powerful tool for uncovering the important degrees
of freedom of a physical system and the interaction forces between
them.  The spectrum of QCD is very rich: conventional baryons
(nucleons, $\Delta$, $\Lambda$, $\Xi$, $\Omega$, {\it etc.}) and mesons
($\pi$, $K$, $\rho$, {\it etc.}) have been known for nearly half a
century, but other, higher-lying exotic states, such as
glueballs, hybrid mesons and hybrid baryons bound by an excited gluon
field, and `multi-quark' states, consisting predominantly of four or five
quarks in the case of mesons and baryons respectively, have proved
more elusive, partly because our theoretical understanding of such
states is insufficient, making their identification difficult.

Interest in excited baryon resonances in particular has been sparked
by experiments dedicated to mapping out the $N^\ast$ spectrum in Hall
B at the Thomas Jefferson National Accelerator Facility (JLab).
Much of our current understanding of conventional and excited hadron
resonances comes from QCD-inspired phenomenological models.  For
conventional baryons, the extensive calculations by Isgur, Karl, and
Capstick within a non-relativistic quark
model\cite{Isgur:1977ef,Isgur:1978xj,Capstick:1986bm} remain
influential.  However, there are a growing number of resonances which
cannot be easily accommodated within quark models.  States bound by an
excited gluon field, such as hybrid mesons and baryons, are still
poorly understood.  The natures of the Roper resonance and the
anomalously light $\Lambda(1405)^-$ remain controversial.  Experiment
shows that the first excited positive-parity spin-$1/2$ baryon lies
below the lowest-lying negative-parity spin-$1/2$ resonance, a fact
which is difficult to reconcile in quark models.  The question of the
so-called ``missing'' baryon resonances is still unresolved: the quark
model predicts many more states\cite{Isgur:1978xj,Capstick:1986bm}
than are currently known.  Compared to the large number of
positive-parity states, there are only a few low-lying negative-parity
resonances. A quark-diquark picture of baryons predicts a sparser
spectrum\cite{Oettel:1998bk}.

Given the current intense experimental efforts in spectroscopy for
baryons, the need to predict and understand the baryon spectrum from
first principles is clear; lattice QCD calculations provide the means
of undertaking such \textit{ab initio} studies.  The aim is not merely
to obtain a set of masses for the states, but also to gain insight
into the quark and gluon structure of the states and to understand the
relevant degrees of freedom; this latter aspect will be an important
emphasis of this talk.  The freedom to vary quark masses, numbers of
quark flavors and even the gauge group enables us not only to relate
lattice computations directly to experiment, but also to QCD-inspired
pictures of baryon structure.  

The layout of the remainder of this talk is as follows.  In the next
section we introduce the variational method as a means of extracting
information about the QCD spectrum, and demonstrate the need to
construct a sufficient basis of interpolating operators.
Section~\ref{sec:group} describes the construction of lattice baryon
operators using a group-theory method\cite{Basak:2005aq}; an
alternative approach we have developed is described in
ref.~\cite{Basak:2005ir}. 

\section{HIGHER EXCITED RESONANCES AND THE VARIATIONAL METHOD}
A comprehensive picture of resonances requires that we go beyond a
knowledge of the ground state mass in each channel, and obtain the
masses of the lowest few states of a given quantum number.  This we
can accomplish through the use of the variational
method\cite{Michael:1985ne,Luscher:1990ck}.  Rather than measuring a
single correlator $C(t)$, we determine a matrix of correlators
\[
C_{ij}(t) = \sum_{\vec{x}} \langle  O_i(\vec{x}, t) O^{\dagger}_j
(\vec{0}, 0) \rangle \label{eq:corrs_cons},
\]
where $\{  O_i; i = 1,\dots,N \}$ are a basis of interpolating
operators with given quantum numbers.  We then solve the generalized
eigenvalue equation
\[
C(t) u = \lambda(t, t_0) C(t_0) u
\]
to obtain a set of real (ordered) eigenvalues $\lambda_n(t,t_0)$,
where $\lambda_0 > \lambda_1 >,\dots$.  At large Euclidean times,
these eigenvalues then delineate between the different masses
\[
\lambda_n (t,t_0) \longrightarrow e^{ -M_n (t-t_0)} + O (e^{ - M_{n+1} (t-t_0)}).
\]
The eigenvectors $u$ are orthogonal with metric $C(t_0)$, and a
knowledge of the eigenvectors can yield information about the partonic
structure of the states.  A recent application of this method to the
baryon spectrum using operators constructed using smeared-quark
sources of varying widths is discussed in ref.~\cite{Burch:2004he}.

\section{BARYON OPERATORS AND THE LATTICE}\label{sec:group}
Crucial to the application of variational techniques is the
construction of a basis of operators that have a good overlap with the
lowest-lying states of interest.  These operators should have the
property that they respect the symmetries of the lattice, rather than
being mere discretisations of continuum interpolating operators.  The
LHP Collaboration has developed techniques to enable the
construction of baryon interpolating operators that can easily be
extended to include multi-quark operators, and those with excited
glue\cite{Basak:2005aq,Basak:2005ir}.

\subsection{Interpolating operators and lattice symmetries}
\begin{table*}
\caption{Continuum limit spin identification:
   the number $n_\Lambda^J$ of times that the $\Lambda$ irrep.\ of the
   octahedral point group $O_h$ occurs in the
   (reducible) subduction of the $J$ irrep.\ of $SU(2)$.  The numbers
   for $G_{1u},G_{2u},H_u$ are the same as for $G_{1g},G_{2g},H_g$,
    respectively.
\label{tab:subduction}}
\renewcommand{\arraystretch}{1.5} 
\begin{center}
\begin{tabular}{cccc@{\hspace{2em}}cccc} \hline
  $J$  & $n^J_{G_{1g}}$ & $n^J_{G_{2g}}$ & $n^J_{H_g}$ &
 $J$  & $n^J_{G_{1g}}$ & $n^J_{G_{2g}}$ & $n^J_{H_g}$\\ \hline
  $\frac{1}{2}$  &  $1$ & $0$ & $0$ &$\frac{9}{2} $ &  $1$ & $0$ & $2$ \\
  $\frac{3}{2}$  &  $0$ & $0$ & $1$ &$\frac{11}{2}$ &  $1$ & $1$ & $2$ \\
  $\frac{5}{2} $ &  $0$ & $1$ & $1$ &$\frac{13}{2}$ &  1 & 2 & 2 \\
  $\frac{7}{2} $ &  $1$ & $1$ & $1$ &$\frac{15}{2}$ &  1 & 1 & 3 \\ \hline
\end{tabular}
\end{center}
\end{table*}
States at rest are classified according to their transformation
properties under rotations; in a lattice calculation, such rotations
are restricted to those of the cubic group of the lattice, $O$.  This
group has the following properties:
\begin{itemize}
\item $O$ has 24 elements
\item There are five conjugacy classes, and hence five single-valued
  representations, $A_1, A_2, E, T_1,~\mbox{and}~T_2$, of dimensions
  1,1,2,3 and 3, respectively.
\end{itemize}
The irreducible representations are defined such that, 
under the elements $R$ of $O$, the operators lying in the
  $\Lambda$ irreducible representation (irrep.) transform as
\[
U(R) O^{(\Lambda)}_{\lambda} U(R)^{\dagger} = 
\sum_{\lambda'} O^{(\Lambda)}_{\lambda'} D^{(\Lambda)}_{\lambda'\lambda}(R)
\]
where $\lambda$ is the row of the irrep., $U$
is the unitary matrix effecting the rotation, and $D$ is the
corresponding representation matrix.

The addition of the spatial inversion operator, corresponding to
parity, yields the \textit{Octahedral Group} $O_h$, and the
irreps.\ acquire an additional subscript $g$ or
$u$, denoting positive and negative parity respectively.  In the case
of baryons, we must consider the double-valued, or spinorial,
irreducible representations.  There are three such irreducible
representations, $G_1$, $G_2$ and $H$, of dimensions 2, 2 and 4
respectively, again with $g$ and $u$ labels to denote parity.

The irreducible representations $J$ of the continuum rotation group
$SU(2)$ are reducible under the cubic group $O$; the number of times
$n^J_{\Lambda}$ that the $\Lambda$ irrep.\ of $O_h$ occurs in the
reducible subduction of the $J$ irrep.\ of the continuum SU(2) is
shown in Table~\ref{tab:subduction}.  States with $J > 5/2$ lie in
irreducible representations containing states with lower spins, and
furthermore, for a given $J$, different continuum helicities can
correspond to members of different irreducible representations of
$O_h$. The masses of the components in these distinct irreps.\ will agree
only in the continuum limit.

An implicit assumption in previous lattice studies is the increase in
ground-state masses with increasing spin.  This assumption is not
necessarily realized in nature; in the nucleon sector, the
lowest-lying $J^P = \frac{5}{2}^+$ state is comparable in mass to the
lowest-lying $J^P = \frac{3}{2}^+$ state\cite{PDBook}.  In a lattice
computation, the spin of an energy level in a given channel can only be
identified by an examination of the degeneracies between energies in
different irreps.\ in the approach to the continuum limit.  Thus the
ability to extract several energy levels in each channel, and
therefore the work outlined in this talk, is especially crucial.

\subsection{Operator recipe}
The starting point for the construction of our three-quark operators
is a basis of gauge-invariant terms of the form
\begin{equation}
\Phi^{ABC}_{\alpha \beta\gamma; i j k}
\!\!=\!    \varepsilon_{abc} (\tilde{D}^{(p)}_i\!\tilde{\psi})^A_{a\alpha} 
   (\tilde{D}^{(p)}_j\!\tilde{\psi})^B_{b\beta}
   (\tilde{D}^{(p)}_k\!\tilde{\psi})^C_{c\gamma},\label{eq:gauge_inv_op}
\end{equation}
where $A,B,C$ indicate quark flavor, $a,b,c$ are color indices,
$\alpha,\beta,\gamma$ are Dirac spin indices, $\tilde{\psi}$ indicates
a smeared quark field, and $\tilde{D}^{(p)}_j$ denotes the $p$-link
covariant displacement operator in the $j$-th direction; the quark
fields are smeared using a three-dimensional gauge-covariant
Laplacian.  

The displacements in \eqn{eq:gauge_inv_op} are chosen so as
to span a range of possible spatial configurations of the quarks, and
are listed in Table~\ref{tab:opforms}.  In particular, the
singly-displaced operators mimic a quark-diquark combination, whilst
the doubly- and triply-displaced operators are chosen since they may
favour $\Delta$- and $Y$-flux configurations, respectively.
\begin{table}[h]
\caption[captab]{The six types of three-quark
$\Phi^{ABC}_{\alpha\beta\gamma;\ ijk}$ operators used, where $A,B,C$
indicate the quark flavors, $1\leq \alpha,\beta,\gamma\leq 4$ are
Dirac spin indices, and $-3\leq i,j,k\leq 3$ are displacement
indices. Smeared quark fields are shown by solid circles, line
segments indicate covariant displacements, and each hollow circle
indicates the location of a color $\varepsilon_{abc}$ coupling.  A
displacement index having a zero value indicates no displacement.
\label{tab:opforms}}
\begin{tabular}{cl}
\hline\\[0.1ex] Operator type &  Displacement indices\\[1.0ex] \hline
\raisebox{0mm}{\setlength{\unitlength}{1mm}
\thicklines
\begin{picture}(16,10)
\put(8,6.5){\circle{6}}
\put(7,6){\circle*{2}}
\put(9,6){\circle*{2}}
\put(8,8){\circle*{2}}
\put(0,0){single-site}
\end{picture}}  & \raisebox{3mm}{$i=j=k=0$ }\\ 
\raisebox{0mm}{\setlength{\unitlength}{1mm}
\thicklines
\begin{picture}(23,10)
\put(7,6.2){\circle{5}}
\put(7,5){\circle*{2}}
\put(7,7.3){\circle*{2}}
\put(14,6){\circle*{2}}
\put(9.5,6){\line(1,0){4}}
\put(0,0){singly-displaced}
\end{picture}}  & \raisebox{3mm}{$i=j=0,\ k\neq 0$} \\ 
\raisebox{0mm}{\setlength{\unitlength}{1mm}
\thicklines
\begin{picture}(26,8)
\put(12,5){\circle{3}}
\put(12,5){\circle*{2}}
\put(6,5){\circle*{2}}
\put(18,5){\circle*{2}}
\put(6,5){\line(1,0){4.2}}
\put(18,5){\line(-1,0){4.2}}
\put(-1,0){doubly-displaced-I}
\end{picture}}  & \raisebox{2mm}{$i=0,\ j=-k,\ k\neq 0$} \\ 
\raisebox{0mm}{\setlength{\unitlength}{1mm}
\thicklines
\begin{picture}(20,13)
\put(8,5){\circle{3}}
\put(8,5){\circle*{2}}
\put(8,11){\circle*{2}}
\put(14,5){\circle*{2}}
\put(14,5){\line(-1,0){4.2}}
\put(8,11){\line(0,-1){4.2}}
\put(-5,0){doubly-displaced-L}
\end{picture}}   & \raisebox{4mm}{$i=0,\ \vert j\vert\neq \vert k\vert,
  \ jk\neq 0$}\\ 
\raisebox{0mm}{\setlength{\unitlength}{1mm}
\thicklines
\begin{picture}(20,12)
\put(10,10){\circle{2}}
\put(4,10){\circle*{2}}
\put(16,10){\circle*{2}}
\put(10,4){\circle*{2}}
\put(4,10){\line(1,0){5}}
\put(16,10){\line(-1,0){5}}
\put(10,4){\line(0,1){5}}
\put(-5,0){triply-displaced-T}
\end{picture}}   & \raisebox{4mm}{$i=-j,\ \vert j\vert \neq\vert k\vert,
 \ jk\neq 0$} \\ 
\raisebox{0mm}{\setlength{\unitlength}{1mm}
\thicklines
\begin{picture}(20,12)
\put(10,10){\circle{2}}
\put(6,6){\circle*{2}}
\put(16,10){\circle*{2}}
\put(10,4){\circle*{2}}
\put(6,6){\line(1,1){3.6}}
\put(16,10){\line(-1,0){5}}
\put(10,4){\line(0,1){5}}
\put(-5,0){triply-displaced-O}
\end{picture}}   & \raisebox{4mm}{$\vert i\vert \neq \vert j\vert \neq
  \vert k\vert,\ ijk\neq 0$}\\ \hline
\end{tabular}
\end{table}
These operators are designed to allow a large number of baryon
operators to be constructed using only a small number of quark
propagators, by exploiting the cubic symmetries of the correlators to
reduce the number of quark sources that need to be employed.  

These gauge-invariant operators are now combined into elemental
operators having the appropriate flavor structure; we assume exact
isospin symmetry, and classify operators according to their isospin,
and strangeness (or charm etc.\ where appropriate).  We thus obtain a
set of gauge- and translational-invariant elemental operators
$B^F_i(t) = \sum_{\vec{x}} B^F(\vec{x},t)$ having the appropriate flavor
structure. The final step in our procedure is to apply the group
theoretical projection to yield a set of operators $B_i^{\Lambda\lambda F}$ that transform according to the row $\lambda$ of the $\Lambda$ irreducible representation:
\begin{equation}
B_i^{\Lambda \lambda F}(t) = \frac{d_\Lambda}{g_{O^D_h}} \sum_{R \in
O^D_h} \Gamma^{(\Lambda)}_{\lambda\lambda}(R) U_R B^F_i(t)
U^{\dagger}_R,\label{eq:proj}
\end{equation}
where $O^D_h$ is the double group of $O_h$, $R$ denotes an element of
$O^D_h$, $g_{O^D_h}$ is the number of elements in $O^D_h$, and
$d_{\Lambda}$ is the dimension of the $\Lambda$ irreducible
representation.

\section{IMPLEMENTATION}
The remainder of this talk will provide a status report of our
program to implement a comprehensive study of baryon spectroscopy in
lattice QCD.  We begin by describing the tuning of the smearing
parameters\cite{Basak:2005gi}.

\subsection{Smearing procedure}
It has long been known that the damping of the couplings to
short-wavelength, high-frequency modes is a crucial requirement for
the extraction of the masses of the lowest-lying states in a lattice QCD
calculation\cite{Gusken:1989ad,Allton:1993wc}.  In our study, we
employ a gauge-covariant Gaussian smearing:
\[
 \tilde{\Psi}(x) = \left(1+\frac{\sigma_s^2}{4n_\sigma}
 \Delta\right)^{n_\sigma}\Psi(x),
\]
where
\begin{eqnarray}
\lefteqn{\Delta\Psi(x)=}\nonumber \\
& & \sum_{k=\pm1,\pm2,\pm3} \left(U_k(x)\Psi(x+\hat{k})-\Psi(x)\right).
\end{eqnarray}
There are two tunable parameters, the smearing radius $\sigma_s$ and
the number of iterations $n_{\sigma}$.

The operators introduced in \eqn{eq:gauge_inv_op} in general involve
gauge-covariant displacements.  In order to reduce the statistical
fluctuations in the gauge fields, and possibly to further damp out the
high-frequency couplings, once can smear the link variables, both in
the Laplacian and in the displacement operators.  A widely adopted
procedure for so doing is that proposed by the APE
collaboration\cite{Albanese:1987ds}, which involves a projection back
onto $SU(3)$.  We instead adopt the analytic ``stout'' smearing
prescription proposed in ref.~\cite{Morningstar:2003gk}, defined by
the iterative procedure
\begin{eqnarray}
 U^{(n+1)}_k(x)&=&\exp\left(i\rho\Theta^{(n)}_\mu(x)\right)U^{(n)}_k(x),\\
\Theta_k(x) &=
&\frac{i}{2}\left(\Omega^\dagger_k(x)-\Omega_k(x)\right)- \nonumber \\
& &
\frac{i}{2N}\mbox{Tr}\left(\Omega^\dagger_k(x)-\Omega_k(x)\right)\\
\Omega_k(x)&=&C_k(x) U^\dagger_k(x)\\
C_k(x) & = & \displaystyle\sum_{i\neq
k}\left(U_i(x)U_k(x+\hat{\imath})U^\dagger_i(x+\hat{k})
\right.  \nonumber \\ 
& & \hspace{-1.3cm}\left. + U^\dagger_i(x-\hat{\imath}) U_k(x-\hat{\imath})
U_i(x-\hat{\imath}+\hat{k}) \right).
\end{eqnarray}
Here there are again two tunable parameters, the number of iterations
$n_\rho$, and the staple weight $\rho$.

\begin{figure*}[p]
\includegraphics[width=6.0in, bb=0 0 567 559]{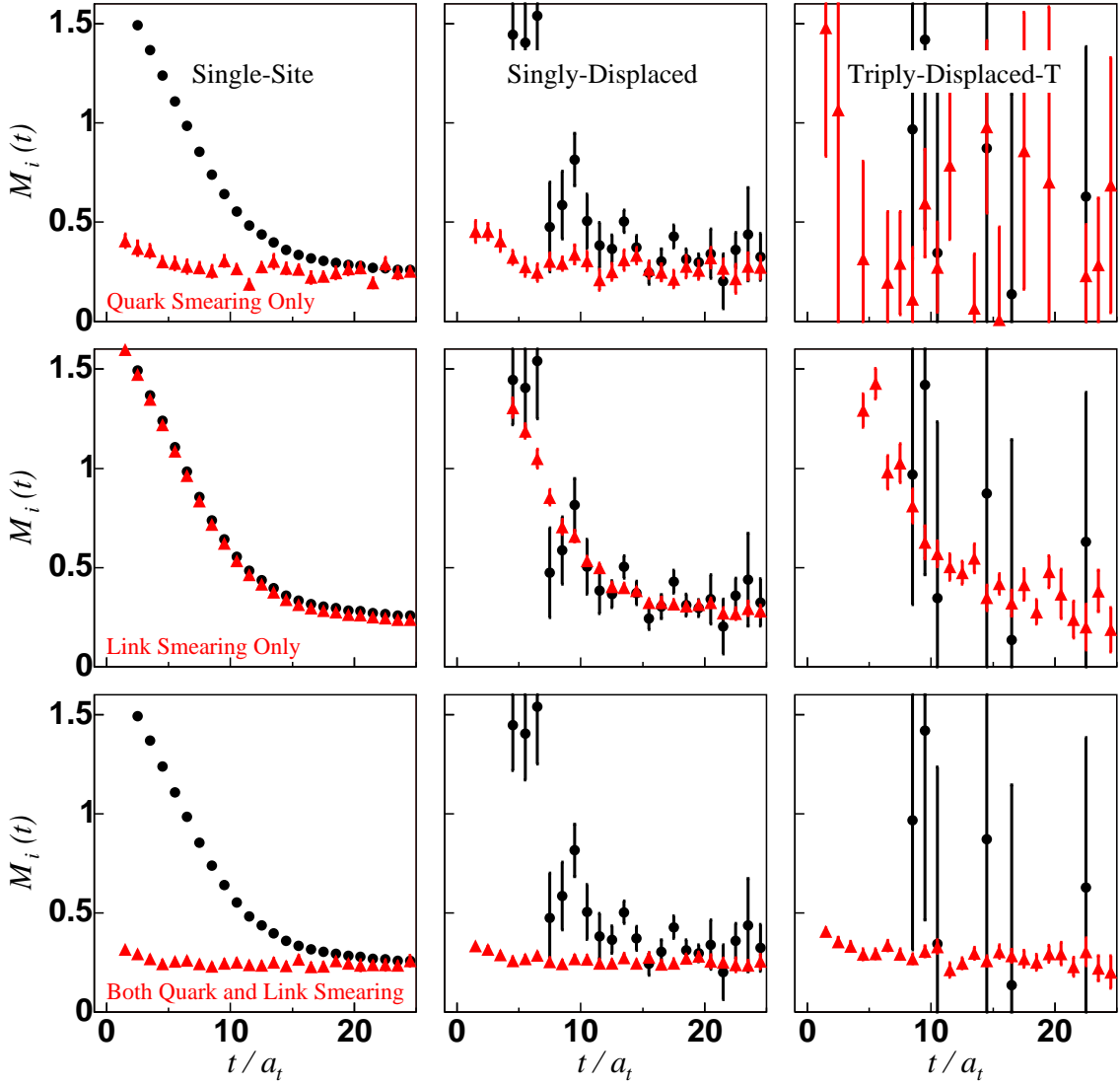}
\caption{Effective masses $M(t)$ for unsmeared (circles) and smeared
(triangles) operators $O_{SS},\ O_{SD},\ O_{TDT}$.  Top row: only
quark-field smearing $n_\sigma=32,\ \sigma_s=4.0$ is used. Middle row:
only link-variable smearing $n_\rho=16,\ n_\rho\rho=2.5$ is applied.
Bottom row: both quark and link smearing $n_\sigma=32,\ \sigma_s=4.0,
\ n_\rho=16,\ n_\rho\rho=2.5$ are used.  Results are obtained on 50
quenched configurations on a $12^3\times 48$ anisotropic lattice using
the Wilson action with $a_s \sim 0.1$ fm, $a_s/a_t =
3.0$\cite{Klassen:1998ua}, as described in the
text.\label{fig:meff-smear}}
\end{figure*}
For our tests of the efficacy of quark-field and gauge-link smearing,
we employed 50 configurations of a quenched, anisotropic $12^3\times
48$ lattice, using the standard Wilson gauge and fermion actions.  The
spatial lattice spacing, determined from the string tension, is $a_s
\simeq 0.1~{\rm fm}$, with a renormalized anisotropy $a_s/a_t = 3$;
the quark mass was chosen so that $m_{\pi} \simeq 700~{\rm MeV}$.
Correlators were computed corresponding to the single-site,
singly-displaced and triply-displaced-T operators of
Table~\ref{tab:opforms}; for the single-site operator, a projection
onto $G_{1g}$ was performed, whilst for the remaining operators a
single Dirac component was chosen with no attempt to project onto an
irreducible representation.

The quark-field smearing parameters $n_{\sigma}$ and $\sigma$ were
determined by requiring that the effective mass for these operators
reach a plateau as close to the source as possible.  The gauge-link
smearing parameters were tuned so as to minimize the noise in the
effective masses; for these lattices we found optimal smearing
parameters $n_\rho \rho = 2.5$ with $n_\rho = 16$.  The use of
gauge-link smearing has only a small effect on the mean values of the
baryon effective masses.  However, we found a dramatic reduction in
the statistical variance in the singly-displaced and triply-displaced
operators, as demonstrated in Figure~\ref{fig:meff-smear}.

\subsection{Group theory projections}
\begin{figure*}[ht]
\includegraphics*[width=6.0in, bb=0 0 567 207]{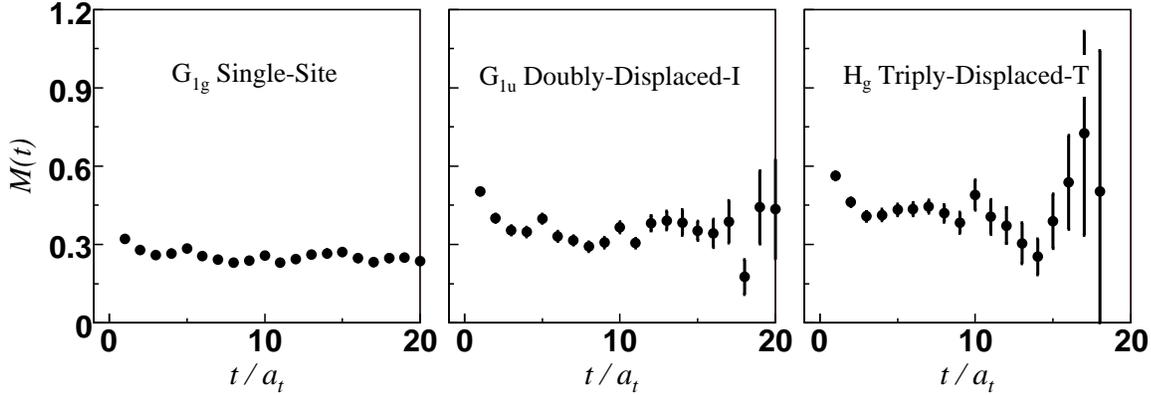}
\caption{Effective masses for three selected nucleon operators:
a single-site operator in the $I = 1/2$ $G_{1g}$ channel (left), a doubly-displaced-I
operator in the $G_{1u}$ channel (center), and a triply-displaced-T
operator in the $H_g$ channel.  The smearing parameters used
were $n_\sigma=32,\ \sigma_s=4.0,\ n_\rho=16,\ n_\rho\rho=2.5$.
\label{fig:meff-irrep}}
\end{figure*}
Having demonstrated the efficacy both of link- and quark-smearing in
isolating the ground-state energies in the correlators, we now proceed
to apply the projection formula of \eqn{eq:proj}.  Effective masses
obtained from three selected nucleon operators in the $G_{1g}$,
$G_{1u}$ and $H_g$ channels, computed using the single-site,
doubly-displaced-I and triply-displaced-T elemental operators, are
shown in Figure~\ref{fig:meff-irrep}.

\begin{figure}
\includegraphics[width=70mm]{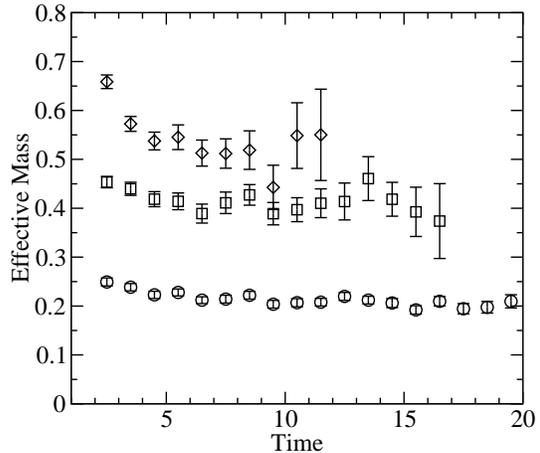}
\caption{The lowest-lying effective masses in
  the $G_{1g}$ channel obtained using the variational
  method\protect\cite{Basak:2004hr}.\label{fig:g1g}}
\end{figure}

\begin{figure}
\includegraphics[width=70mm]{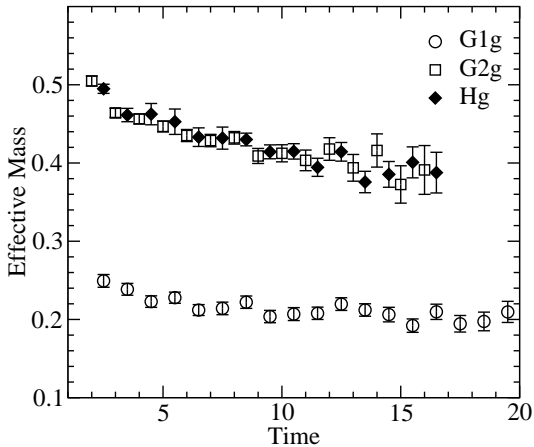} 
\caption{The
 ground-state effective masses in the $G_{1g}$, $G_{2g}$ and $H_g$
 irreducible representations.\label{fig:g2g}}
\end{figure}

An exploratory study of the nucleon spectrum using the Clebsch-Gordon
method\cite{Basak:2005ir} was performed in ref.~\cite{Basak:2004hr}.
Here a larger $16^3 \times 64$ lattice was employed using around 300
configurations with a quark mass $m_{\pi} \simeq 500~{\rm MeV}$.  Such
an ensemble was sufficient to enable the application of the
variational method to extract the lowest-lying eigenvalues of the
$G_{1g}$ correlator matrix, as shown as Figure~\ref{fig:g1g}.
Furthermore, a plateau was found in the $G_{2g}$ irrep.\, one
accessible only through the use of displaced, rather than single-site,
operators.  The effective masses of the ground states in each of the
positive-parity $G_{1g}, H_g$ and $G_{2g}$ irreducible representations
are shown in Figure~\ref{fig:g2g}; the apparent coincidence between
the $H_g$ and $G_{2g}$ effective masses suggests that the
identification of the lowest-lying $H_g$ nucleon state with spin-$3/2$
is somewhat premature, and enforces the need for the program outlined
in this talk.

\section{CONCLUSIONS}
We have outlined a program to study the resonance spectrum in lattice
QCD.  The use of the variational method and the need to isolate
several energy levels in each channel require a sufficiently broad
basis of operators.  Having developed suitable group-theory methods to
project operators onto the irreducible representations of the cubic
group, and having examined the efficacy of both quark- and
gauge-link-smearing, we are now identifying a more limited set of
operators that we will employ in a large-scale study of the hadron
spectrum.  Our methods are applicable not only to baryons, but also to
mesons, to states with excited glue, and to multi-quark and
multi-hadron states.  Only by performing such a program can we hope to
identify the states of QCD, and in particular their spins and
parities, in the continuum limit.

\section*{Acknowledgments}
This work was supported by the U.S.~National Science Foundation
through grants PHY-0099450 and PHY-0300065, and by the U.S.~Department
of Energy under contracts DE-AC05-84ER40150 and DE-FG02-93ER-40762.
Computations were performed using the \textit{Chroma} software
package\cite{Edwards:2004sx}.

\bibliography{writeup}
\end{document}